# Transfer Learning with Edge Attention for Prostate MRI Segmentation

Xiangxiang Qin

East China University of Science and Technology

**Abstract.** Prostate cancer is one of the common diseases in men, and it is the most common malignant tumor in developed countries. Studies have shown that the male prostate incidence rate is as high as 2.5% to 16%, Currently, the incidence of prostate cancer in Asia is lower than that in the West, but it is increasing rapidly. If prostate cancer can be found as early as possible and treated in time, it will have a high survival rate. Therefore, it is of great significance for the diagnosis and treatment of prostate cancer. In this paper, we propose a transfer learning method based on deep neural network for prostate MRI segmentation. In addition, we design a multi-level edge attention module using wavelet decomposition to overcome the difficulty of ambiguous boundary in prostate MRI segmentation tasks. The prostate images were provided by MICCAI Grand Challenge-Prostate MR Image Segmentation 2012 (PROMISE 12) challenge dataset.

## 1 Introduction

With the increase of population and the change of living habits, the incidence and mortality of prostate cancer have increased significantly in recent years. According to the latest statistics of American Cancer Society (ACS) in 2012, the incidence of prostate cancer in European and American countries ranks the first among all male malignant tumors, and the mortality caused by prostate cancer ranks the second, second only to lung cancer. In 2015, academician He Jie of China Cancer Center published the latest incidence and mortality data of prostate cancer in China [1]. The data shows that the mortality of prostate cancer in China increased by 5.5% in 2000-2011, 12.6% in 2000-2005 and 4.7% in 2005-2011. Clinical experience shows that if prostate cancer can be found as early as possible and treated in time, it will have a high survival rate. Therefore, it is of great significance for the diagnosis and treatment of prostate cancer.

Since deep learning has the excellent ability to learn multilevel feature representation from data, it has a very high position in various applications in recent years. Many researchers have employed CNNs in automated prostate segmentation, the classic example is U-Net [2], followed a number of approaches are based on U-Net, which has a high performance in semantic segmentation. Oktay et al [3] proposed an attention gate (AG) model which enable the U-Net model automatically learns to focus on target structures of varying shapes and sizes. Alom et al [4] combined residual convolutional unit and recurrent convolutional block, proposed the recurrent residual convolutional units (RRCU), They replaced the unit with the convolution layers in U-Net model. The improved structure is called R2U-Net. Milletari et al [5] presented a novel 3D segmentation approach: V-Net, they train the network with a dice coefficient loss which can deal with situations where there is a strong imbalance between the number of foreground and background voxels. Zhu et al [6] proposed a boundary-weight transfer learning based domain adaptive neural network (BOWDA-Net), won the first place in the PROMISE12 challenge [7] in 2018. Chen et al [8] proposed a Med3D Network, which is a series of pre-trained models developed specifically for 3D medical imaging in deep learning applications. Jia et al [9] combined a 3D segmentation decoder and a 2D boundary decoder, proposed a Hybrid Discriminative Network (HD-Net), is the state of the art method on PROMISE12 dataset.

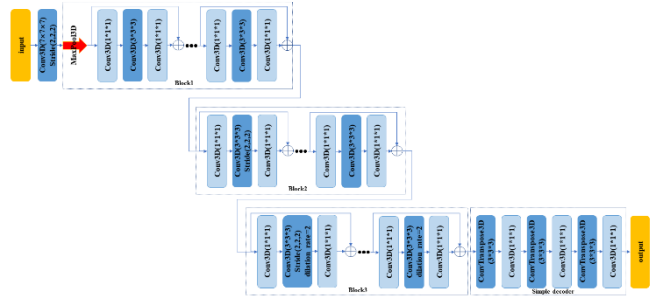

**Fig. 1.** An overview of pre-training encoder architecture

In this paper, we are inspired by Med3D Network, trained an 3D ResNet101 [10] on our datasets as a pre-trained encoder for transfer learning which can accelerate model convergence and reduce the model dependence on data volume. In addition, we propose a novel 3D decoder with multi-level edge attention module (MLEAM), which can enhance and monitor the edge information of feature maps on multiple scales. We trained the pre-trained model on the PROMISE12 dataset and submitted the results of the test set to the challenge official website.

## 2 Method

### 2.1 3D Pre-trained Encoder

We pre-trained a 3D encoder similar to the ResNet101 on the medical MRI dataset of the partner hospital and use dilated convolutional layers with rate 2 in the blocks3. In the pre-training process, we only designed a simple decoder to allow the network to focus on training a universal encoder. The pre-trained encoder network framework is shown in Fig. 1. In the experiment of PROMISE12 dataset, we designed a more complex decoder, which combines the multi-level edge attention mechanism (MLEAM) and pyramid attention module (PAM), the specific structure will be introduced in Section 2.2, 2.3 and 2.4.

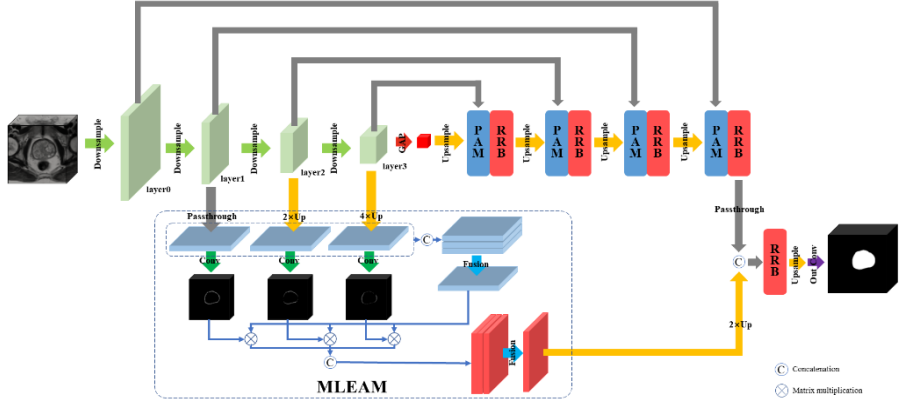

**Fig. 2.** An overview of proposed 3D Multi-Scale Decoder and MLEAM

## 2.2    3D Multi-Scale Decoder

Different from the pre-training process, in the segmentation experiment on PROMISE12 dataset, we designed a decoder (see Fig. 2) that combines multi-scale information and edge information. We use the Residual Refinement Block (RRB) proposed by [9], which replaces the $3 \times 3 \times 3$ convolutions in the residual convolution block with a $3 \times 3 \times 1$ convolution and a $1 \times 1 \times 3$ convolution, the $3 \times 3 \times 1$ convolution helps to capture the 2D features inside the x-y planes, and the $1 \times 1 \times 3$ convolution can focus on between-slices features. The structure of the RRB is given in Fig. 3(a).

## 2.3    Multi-Level Edge Attention Module

One of the major difficulties in 3D prostate segmentation is the ambiguity of the boundary of some regions, which affects the prediction accuracy of the whole 3D voxel. On the last three levels of the designed decoder, we use convolution layers to extract multi-level edge information, and use the attention mechanism to supervise the decoder's capture of edge information. The structure of designed decoder and MLEAM is shown in Fig. 2. We also apple the long skip connection between the corresponding layers of encoder and decoder.

The MLEAM extracts three levels of edge information, which is used to guide the fusion features extracted from the encoder, and fuses the edge feature maps again for the prediction output of the last layer. The process of extracting edge information needs to use the edge map of ground truth for deep supervision.

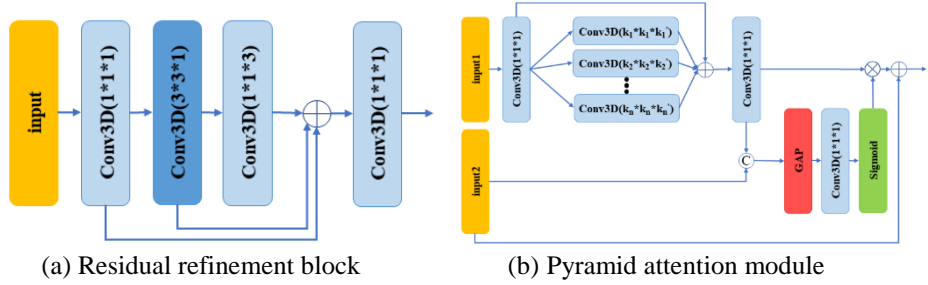

(a) Residual refinement block       (b) Pyramid attention module

**Fig. 3.** Detailed structures of the RRB and PAM

### 2.4 Pyramid Attention Module

The input of each PAM consists of two parts: the output of each encoding layer and the upsampling output of each decoding layer. The structure of PAM is shown in Fig. 3(b).

The encoder output from the long connection passes through multiple convolutional layers of different sizes to generate more discriminative features, Specifically, we use large kernel convolution in the PAM, which can fuse the local and global image contents at multiple scales and reduce information loss.

The upsampling output of the encoder is used to weight the output of the pyramid convolution, in order not to discard the information of the encoder feature map, the output of the encoder is added after weighting.

## 3 Implementation

We implemented our proposed method based on the Pytorch framework with one Nvidia Geforce GTX 1080Ti 11GB GPU. In the process of pre-training on our own data, we unify the spatial resolution of each volume to $0.625 \times 0.625 \times 1.5$ mm, and use SGD optimizer with momentum of 0.9, the initial learning rate is 0.01 and decrease by a weight decay of $1.0 \times 10^{-6}$ after each epoch. The loss function in the pre-training process is cross entropy ($L_{ce}$). Let $y$ represents ground truth and $\hat{y}$ be a segmentation result, $L_{ce}$ can be computed as

$$L_{ce} = -\sum_{y} y \log(\hat{y}) + (1 - y) \log(1 - \hat{y}) \tag{1}$$

Our multi-scale 3D encoder-decoder network is trained end-to-end on a dataset of prostate scans in MRI, There are 50 training MRI volumes from PROMISE 12 challenge dataset and all the volumes processed by the network have fixed size of spatial resolution of $0.625 \times 0.625 \times 1.5$ mm. During every training iteration, we fed as input to the network randomly deformed versions of the training images by using dense deformation field obtained through a $2 \times 2 \times 2$ grid of control-points and B-spline interpolation, we also randomly cropped sub-volumes in the size of $96 \times 96 \times 32$ voxels from the training data, these data augmentations were performed online, it can reduce the potential overfitting caused by limited training images. The mini-batches used in our implementation is 16. Due to the addition of deep edge information supervision, the whole loss function is

$$L_{total} = L_{dice} + \sum_{i=1}^{n} w^i L_{edge}^i \tag{2}$$

Where $L_{dice}$ is calculated from the final segmentation result and ground truth, $w^i$ and $L_{edge}^i$ represent the weight and the 2-norm of the feature map boundary and ground truth boundary of $i$-th layer, we empirically set the weights $w^{i=1,2,3}$ as 0.5, 0.8, 1.0. The calculation formula of $L_{dice}$ and $L_{dege}$ is

$$L_{dice} = 1 - \frac{2 \sum_{i=1}^{N} y_i \hat{y}_i}{\sum_{i=1}^{N} y_i^2 + \sum_{i=1}^{N} \hat{y}_i^2} \tag{3}$$

$$L_{edge} = \frac{1}{N} \left\| M_y - M_{\hat{y}} \right\|^2 \tag{4}$$

Where $M_y$ and $M_{\hat{y}}$ represent the edge map of ground truth and the encoder output.

We trained the network using Adam optimizer with betas of (0.9, 0.999), the initial learning rate is 0.001 which decrease by one order of magnitude every 2K iterations. In the inference phase, for each MR image, we use overlapping sliding windows to crop sub-volumes and used the average of the probability maps of these sub-volumes to get the whole volume prediction. The sub-volume size was $96 \times 96 \times 32$ and the stride was $24 \times 24 \times 8$.

## 4 Experiments and Result

We are submitting the results to PROMISE12, the overall experimental details and results will be given later.